# 7 Tesla multimodal MRI dataset of *ex-vivo* human brain

Qinfeng Zhu[1*], Sihui Li[1*], Zuozhen Cao[1], Yao Shen[1], Haoan Xu[1], Guojun Xu[1], Haotian Li[1], Keqing Zhu[2], Zhiyong Zhao[3+], Jing Zhang[2,4+], Dan Wu[1+]

**Affiliations**
1 Key Laboratory for Biomedical Engineering of Ministry of Education, Department of Biomedical Engineering, College of Biomedical Engineering & Instrument Science, Zhejiang University, Hangzhou, China.
2 China Brain Bank and Department of Neurology in Second Affiliated Hospital, Key Laboratory of Medical Neurobiology of Zhejiang Province, and Department of Neurobiology, Zhejiang University School of Medicine, Hangzhou, China.
3. Children's Hospital, Zhejiang University School of Medicine, National Clinical Research Center for Child Health, Hangzhou, China.
4 Department of Pathology, The First Affiliated Hospital and School of Medicine, Zhejiang University, Hangzhou, China.

* These authors contributed equally to this manuscript
Correspond to Dan Wu (danwu.bme@zju.edu.cn)

**Abstract**

*Ex-vivo* MRI offers invaluable insights into the complexity of the human brain, enabling high-resolution anatomical delineation and integration with histopathology, and thus, contributes to both basic and clinical studies on normal and pathological brains. However, *ex-vivo* MRI is challenging in sample preparation, acquisition, and data analysis, and existing *ex-vivo* MRI datasets are often single image modality and lack of ethnic diversity. In our study, we aimed to address these limitations by constructing a comprehensive multimodal MRI database acquired from six *ex-vivo* Chinese human brains. This database included structural MRI, high-angular resolution diffusion MRI, quantitative susceptibility mapping, and quantitative T1 and T2 maps, which enabled multifaceted depiction of brain microstructure and connectivity. Furthermore, we generated population-averaged multimodal templates and the segmentation labels to facilitate analysis of *ex-vivo* brain MRI. This public database offers a collection of high-resolution and multi-parametric *ex-vivo* human brain MRI and filled the gap of lacking Asian brain samples in existing databases.

**Background & Summary**

Magnetic resonance imaging (MRI) provides a non-invasive way to investigate the brain anatomy and function with versatile contrasts. Given the intricated and complex architecture of the human brain, high resolution is desired for detailed delineation of the brain anatomy. For instance, the human cortex is microscopically defined into six different cytoarchitectonic layers, ranging from about 0.2 mm to 0.8 mm[1], based on differences in cell types, packing density, or myelinated neurites [2]. The subfields in hippocampus is also divided into functionally distinct layers, with the stratum granular (SG) layer of the dentate gyrus (DG) being only 0.2 mm thick [3]. Due to limited spatial resolution of *in-vivo* MRI (typically > millimeter), it is difficult to distinguish the cortical layers [4] or hippocampal layers. Although high-resolution *in-vivo* MRI has been shown feasible with high-performance hardware[5] and advanced pulse sequences [6,7], it still takes a long scan time and remains challenging in clinical settings. *Ex-vivo* MRI provide the opportunity to reach ultra-high-resolution over *in-vivo* scans due to unlimited scan time, absence of motion, and specialized hardware setups [8,9].
Several groups have actively working in the field of *ex-vivo* human brain MRI, and the reported the advantages of *ex-vivo* human brain MRI. Weigel et al. [10] developed a 3D ultra-high-resolution imaging approach for *ex-vivo* whole-brain MRI of multiple sclerosis patients on a standard clinical 3 T MRI system. The reconstructed T2 weighted images at 160 μm isotropic resolution allowed visualization of sub-millimetric lesions in the different cortical layers and in the cerebellar cortex. Zhao et al. [11] performed the diffusion MRI (dMRI) of the anterior hippocampal blocks on an ultra-high field of 14.1 T to assess Alzheimer's disease-induced changes in hippocampal microstructure. This study observed seven hippocampal layers [12] at 100 μm isotropic resolution, which was not possible in *in-vivo* scans. Furthermore, the utilization of sub-millimeter *ex-vivo* dMRI has been shown to decrease the occurrence of erroneous outcomes in brain fiber tracking [13,14], thereby facilitating the identification of more precise structural connections.

Another advantage of *ex vivo* MRI lies in that it enables cross-modal correlation between *ex-vivo* MRI and optically-clear histology. Currently, there are studies that have achieved systematic correlation and co-registration of histological and MRI images of the adult brain [15-17]. The integration of *ex-vivo* MRI with histologic staining informs our biological understanding of diseased brains across microscopic and mesoscopic scales [11,18,19]. For multiple sclerosis (MS), Alzheimer's disease (AD) and Parkinson's disease, their pathogenesis is related to changes in iron homeostasis and myelin production. These two major biomarkers were demonstrated to be detectable in *ex-vivo* specimen by quantitative MRI (qMRI) [20], such as T2* map [21,22] and quantitative susceptibility mapping (QSM) [23], by quantitative comparison with Luxol Fast Blue (LFB) stain and iron stain. In addition, disease-induced abnormal interbrain connectivity can be detected by dMRI [3,24,25], and various diffusion-related models such as Diffusion Basis Spectrum Imaging (DBSI) [26], Neurite Exchange Imaging (NEXI) [27], etc. had been proposed to explain microstructural changes. The above mentioned studies offered proper biological interpretation of *in-vivo* MRI findings by exploring correlations between *ex-vivo* MRI and histological images [15].

Despite the advantages of *ex-vivo* MRI, the public databases are relatively rare. Existing *ex-vivo* databases, such as BigBrain [16] and Julich Brain [17], focused on immunohistological atlases, and utilized MRI as an anatomical reference for section localization and brain region segmentation. Databases such as the Allen Human Brain Atlas (AHBA) [28] and BrainSpan (www.brainspan.org) provided transcriptome atlas of the human brain, and used multimodal MRI data to assist the localization of tissue blocks. The Digital Brain Bank (DBB)[13] disclosed a ultra-high-resolution *ex-vivo* human brain data with several imaging contrasts acquired at 7Tesla along with Polarised Light Imaging (PLI). DBB included structural MRI, T1 and T2 map, as well as dMRI with three different resolutions (500μm, 1mm, 2mm isotropic). However, the MRI data in the existing open-source *ex-vivo* databases are limited to a single individual of selected age and gender, which is not representative of population. Moreover, advanced image modalities, such as quantitative susceptibility mapping (QSM) that is known to be sensitive to myelin and iron deposition, are not incorporated in the existing studies. In addition, all previous *ex-vivo* MRI databases were exclusively of non-Asian origin, while studies have demonstrated a considerable difference between brains from different ethnic origins. In morphological studies, significant differences were found between Asians and Caucasians in the thickness, volume, and surface area of most cortical brain regions [29,30]. Differences in gene expression between races have also been reported [31-33].

In this study, we published a multimodal MRI database collected from six *ex-vivo* Chinese human brains without known neurological diseases, and provided multi-metric templates derived from this database. High-resolution MRI of the right hemisphere was acquired at a 7T scanner, covering 3D T2-weighted MRI, quantitative T1 and T2 maps, 3D dMRI for diffusion tensor metrics and tissue microstructural properties using DBSI (Diffusion Basis Spectrum Imaging) model [26], and 3D QSM to obtain positive and negative susceptibility maps and T2* maps. Furthermore, we provided a pipeline to generate population-average templates for different modalities and quantitative maps, which may serve as an unbiased reference for *ex-vivo* studies.

**Methods**

*Specimen preparation*

The study used six adult *ex-vivo* hemisphere specimens obtained from the National Health and Disease Human Brain Tissue Repository [34]. The donors or their next-of-kin who were authorized by donors or by law signed the informed consent form for a clinical autopsy and for donation of her brain for research, as part of a protocol approved by our Institutional Review Board. The neuropathologic examination of these donors revealed no significant abnormalities. Table 1 presents the clinical information of the specimens utilized in this paper. After the autopsy, the specimens were fixed in a phosphate buffer (pH 7.2-7.4) configured with 6% paraformaldehyde (PFA) for 28 days.

Table. 1 Summary of brain specimens

| Specimen No. | gender | Age | Clinical diagnosis | Brain weight (g) | Autopsy delay time | Fixed duration (day) |
|---|---|---|---|---|---|---|

|   |   |   |   |   | (hour) |   |
|---|---|---|---|---|---|---|
| 1 | Male | 63 | Stage III malignant tumor of the esophagus; Secondary malignant tumor of the lymph nodes | 1065 | 16 | 53.3 |
| 2 | Male | 69 | Shock; Respiratory failure | 1249 | 15 | 35.5 |
| 3 | Male | 58 | Diffuse large B cell lymph nodes | 1396 | 17 | 72.2 |
| 4 | Male | 64 | Malignant tumor of gallbladder (adenocarcinoma) | 1279 | 6 | 27.4 |
| 5 | Male | 54 | Malignant tumors of the chest (bone, lung, thymus, lymph nodes) | 1429 | 5 | 37.5 |
| 6 | Female | 48 | Gastric tumor; Malignant tumor of the ovary | 1162 | 11 | 30.6 |

Figure 1 illustrates the preparation of one *ex-vivo* specimen (Fig. 1a). We first rinsed and soaked the specimen in saline for two days (Fig. 1b). Second, we dried the cortical sulcus as well as the ventricles with absorbent paper, and transferred the specimen to a homemade container. We heavily padded both the brain in the container with sponges to minimize mechanical coupling between the specimen and the MRI system during data acquisition [35] (Fig. 1c). Next, we filled the container with Fomblin to cover the entire specimen for susceptibility matching, and placed the container under negative pressure of 0.095 MPa for 2 days to remove the air (Fig. 1d). During this period, detritus and water may float on the surface of the container, which need to be skimmed off to eliminate interference signals. Finally, we sealed the entire container.

Note that containers were homemade to adapt to the size of the *ex-vivo* brain, using 3D printing. The walls of the container were made of photosensitive resin material, while the container cover was made of acrylic material. Both materials do not produce electromagnetic signals. The cover of the container was designed with three layers of acrylic plates. The innermost two layers of acrylic plates had holes designed to facilitate the outflow of bubbles and water from the specimens and the outermost layer sealed the entire container with resin bolts.

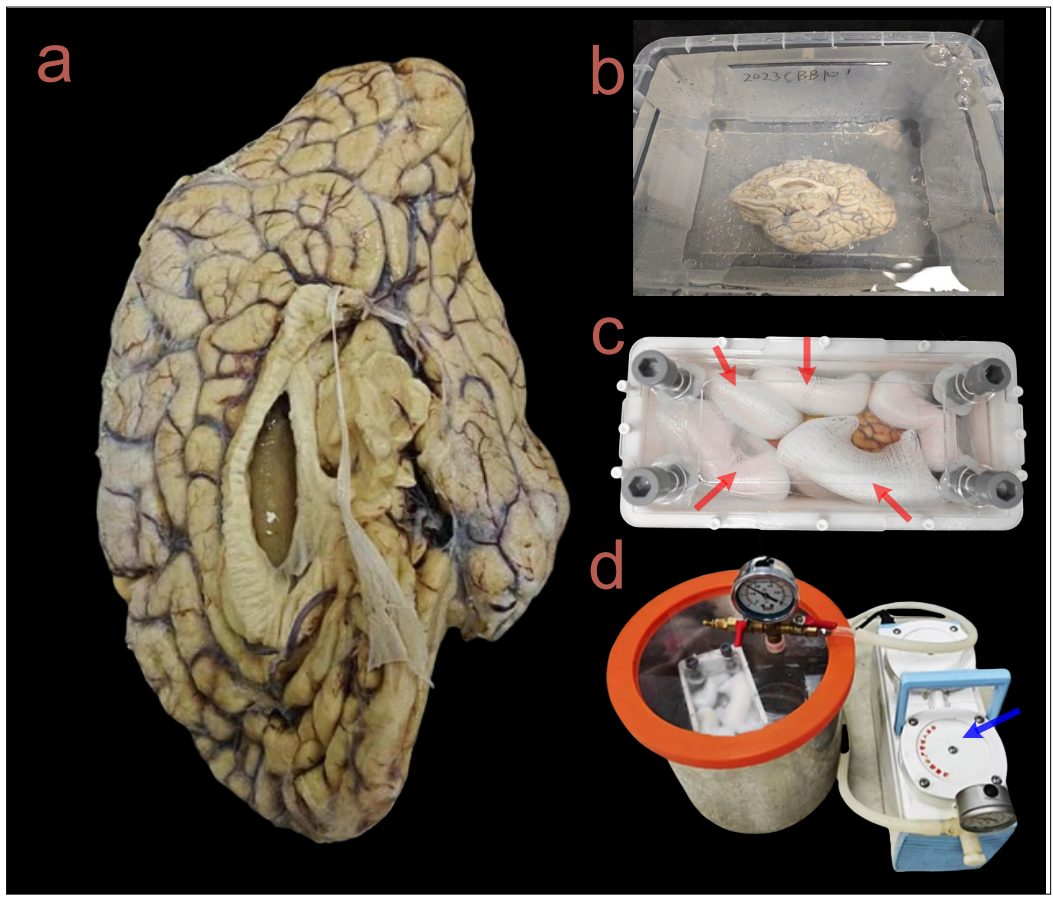

**Fig. 1** Preparation of an *ex-vivo* human brain specimen. (a) A gross look of the *ex-vivo* specimen of the right hemisphere of a human brain. (b) The formalin-fixed specimen was soaked in saline before scanning. (c) The specimen was immobilized using a 3D-printed container. To prevent specimen movement during scanning, sponges (denoted by the red arrows) were placed around the specimen. (d) Air bubbles in the *ex-vivo* brain was removed by a vacuum pump (denoted by the blue arrow).

**Data acquisition**

All images were acquired on a 7T MRI scanner (MAGNETOM 7T, Siemens Healthcare, Erlangen, Germany) equipped with a whole-body gradient set (70 mT/m and 200 T/m/s) and a 32 channel head coil. Multi-modal MRI including 3D T2w, 3D high-angular resolution diffusion MRI (HARDI), QSM, and quantitative T1 and T2 maps were performed. 3D T2w was acquired with a variable flip angle 3D turbo spin-echo sequence at a resolution of 0.5 x 0.5 x 0.6 mm³. 3D HARDI was acquired with a

3D diffusion-weighted steady-state free precession (SSFP) sequence at 0.8 isotropic resolution with 60 diffusion directions. A relatively high b-value of 6000 mm/s$^2$ was used considering the much lower diffusivity in *ex-vivo* specimen than *in-vivo* brains [36]. T1 map was acquired with a 2D inversion recovery spin-echo sequence at a resolution of 0.6 x 0.6 x 1.2 mm$^3$ with 4 inversion times, and T2 map was acquired with a 2D spin-echo sequence at same resolution with 5 echo times. QSM was acquired with a 3D multi-echo gradient-recalled echo sequence at a resolution of 0.6 x 0.6 x 1.0 mm$^3$ with 12 echo times. Detailed acquisition parameters were listed in Table 2 and the multimodal data were illustrated in Fig. 2.

**Table. 2** Summary of sequence parameters

| Model | Sequence | TE (ms) | TR (ms) | FOV (mm²) | Number of slices | Voxel size (mm³) | TI (ms) | Diffusion scheme | Scan time (min) |
|---|---|---|---|---|---|---|---|---|---|
| 3D T2w | Variable flip angle 3D turbo spin-echo | 24 | 12510 | 180 x 112 | 128 | 0.5 x 0.5 x 0.6 | | | 0:38 |
| HARDI | 3D diffusion-weighted steady-state free precession | 21 | 29 | 180 x 113 | 96 | 0.8 x 0.8 x 0.8 | | 20 b0s and 60 directions at b = 6000 mm²/s | 5:06 |
| T1 Map | 2D inversion recovery spin-echo | 14 | 5660 | 180 x 112 | 18 | 0.6 x 0.6 x 1.2 | 4 TIs at 50/200/400/800 | | 0:57 |
| T2 Map | 2D spin-echo | 5 TEs at 8.8/18/28/38/48 | 5000 | 180 x 112 | 18 | 0.6 x 0.6 x 1.2 | | | 1:02 |
| QSM | 3D multi-echo gradient-recalled echo | 12 TEs at 4.0/8.1/12.2/16.4/20.5/24.6/28.8/32.9/37.1/41.2/45.2/49.3 | 100 | 180 x 112 | 72 | 0.6 x 0.6 x 1.0 | | | 3:10 |

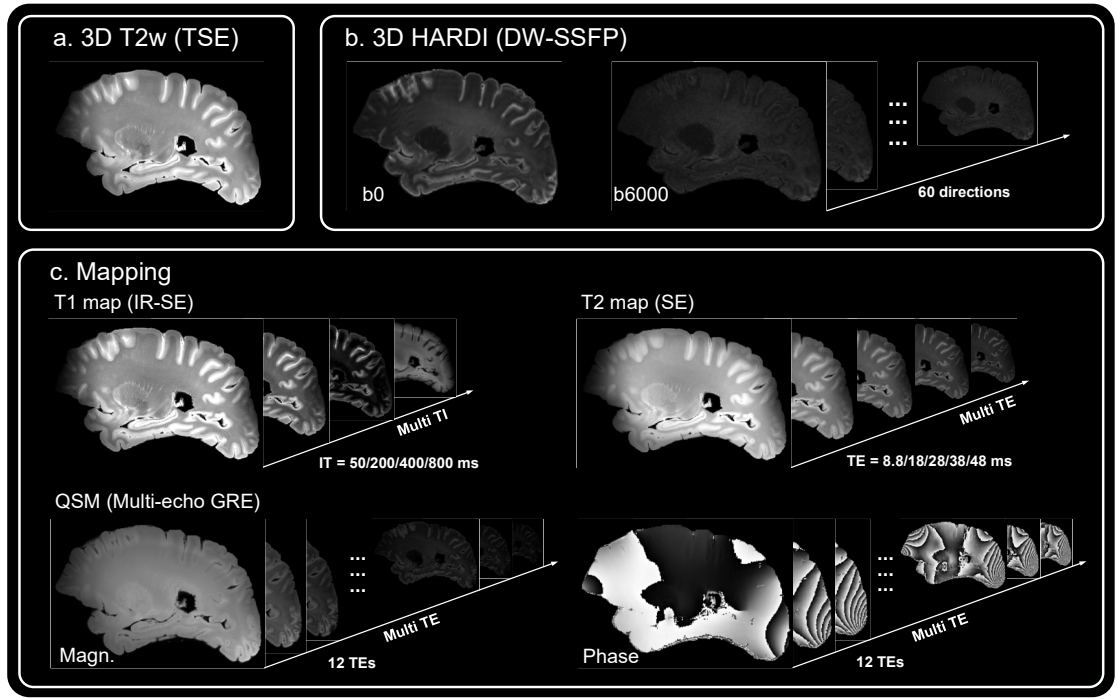

**Fig. 2** Raw data of the acquired multimodal MRI. (a) Sagittal view of the 3D T2w MRI using T2-weighted Turbo Spin Echo (TSE) sequence. (b) HARDI obtained by a 3D DW-SSFP (diffusion-weighted steady-state free precession) sequence. The protocol included 20 non-diffusion scans and 60 diffusion directions at b-value of 6000 s/mm$^2$. (c) Quantitative mapping data including T1 mapping (images at different inversion times), T2 (images at different spin-echo times), and QSM (magnitude and phase images acquired at 12 gradient-echo times).

*Parameter fitting*

The HARDI data were preprocessed using the pipeline in MRtrix 3.0 (version 3.0_RC3-135-g2b8e7d0c) [37]. Denoising and Gibbs artifact removal were performed on the original images, followed by motion and distortion correction. Then, three dMRI processing techniques, namely DTI, DBSI [26], and constrained spherical deconvolution (CSD) to obtain fiber orientation density function (FOD) [38], were performed on preprocessed DWI data. Specifically, the tensor model was fitted in MRtrix to obtain the mean apparent diffusion coefficient (MD), fractional anisotropy (FA), axial diffusivity (AD), and radial diffusivity (RD). The DBSI data was analyzed using a specialized toolbox [26], which generated fiber fraction ($f_{fiber}$), hindered isotropic fraction

($f_{hindered}$) that represented extracellular diffusion, restricted isotropic fraction ($f_{restricted}$) that represented intracellular diffusion, and free isotropic fraction ($f_{free}$) that represented cerebrospinal fluid (CSF). The fitting process involved defining diffusivity ranges for each component, with 0 to 0.3x10$^{-3}$ mm²/s for intracellular component, 0.3 to 3 x10$^{-3}$ mm²/s for the hindered water component, and the cerebrospinal fluid (CSF) component ranged from 2 x10$^{-3}$ mm²/s to positive infinity. The FOD was obtained using CSD in MRtrix, which was used to quantify fiber density (FD).

The T1- and T2-relaxation time maps were fitted separately based on $S(TI) = M_0 + ae^{(-TI/T_1)}$ and $S(TE) = M_0 e^{(-TE/T_2)}$, where $M_0$, $a$, $T_1$, $T_2$ are unknown. T2* maps were obtained from the QSM data by fitting the magnitude images using the same program as for the T2 map fitting.

The QSM was processed using STISuite (version 2.2) [39]. The amplitude and phase maps obtained from the multi-echo GRE were first input to perform 3D phase unwrapping and background phase removal using a Laplace-based method [39]. The apparent magnetic susceptibility from background phase removal was then calculated using the Matlab LSQR linear solver [40]. Finally, the k-space based method was used to calculate the susceptibility maps, including the absolute susceptibility map, positive susceptibility map, and negative susceptibility map.

Fig. 3 displays all the quantitative maps obtained in an *ex-vivo* brain. Compared with *in-vivo*, the T1, T2, and T2* relaxation times, as well as the overall diffusivity, of fixed *ex-vivo* specimens were much lower [36]. It was observed that since cortical gray matter (GM) is rich in iron and white matter (WM) is rich in myelin [41], they show high signals in positive and negative susceptibility maps, respectively. Among the QSM-related maps, it can be found that the external capsule (EC) showed extremely high negative susceptibility, which was consistent with previous reports [42,43,44]. From DTI, it was observed that GM had higher diffusivity than WM, while WM exhibited higher FA. In DBSI, strong fiber components were seen in the fornix and posterior corona radiata, while $f_{hindered}$ and $f_{restricted}$ were high in the cortical GM. The fiber distribution in FOD was consistent with that in $f_{fiber}$ map, and the orientation information of the fibers was use for fiber tracking.

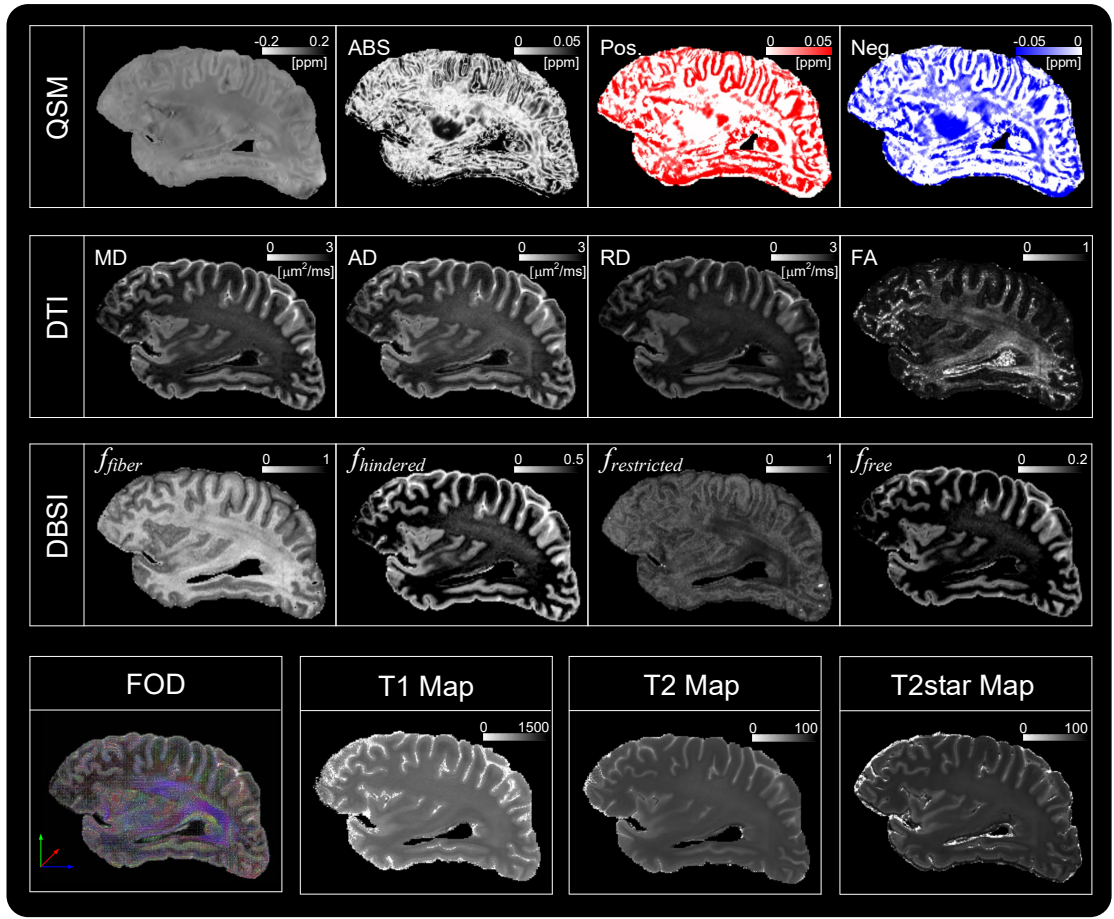

**Fig. 3** Multimodal quantitative maps. QSM involves the generation of absolute susceptibility maps, as well as positive and negative susceptibility maps, obtained from multiple gradient echo data. Three diffusion-related modes, including DTI that gave rise to mean diffusivity (MD), axial diffusivity (AD), radial diffusivity (RD), and fractional anisotropy (FA), DBSI that generated $f_{fiber}$, $f_{hindered}$, $f_{restricted}$, and $f_{free}$ map, and FOD were shown in sagittal views. T1, T2, and T2* relaxation maps were shown in the bottom row.

### *Visualization of WM structures*

Fiber tractography was computed based on the FOD map using the MRtrix software (v.3.0.3) using the iFOD2 algorithm, with 27 seeds in each voxel, min length of 5 mm, max length of 400 mm, step size of 0.2 mm and relative threshold of 0.1. Tractography and seeding were constrained to the brain WM regions, which were used to generate high resolution track density imaging (TDI) at 100 μm resolution. Detained fiber structures can be clearly discriminated on the superresolution TDIs (Fig. 4).

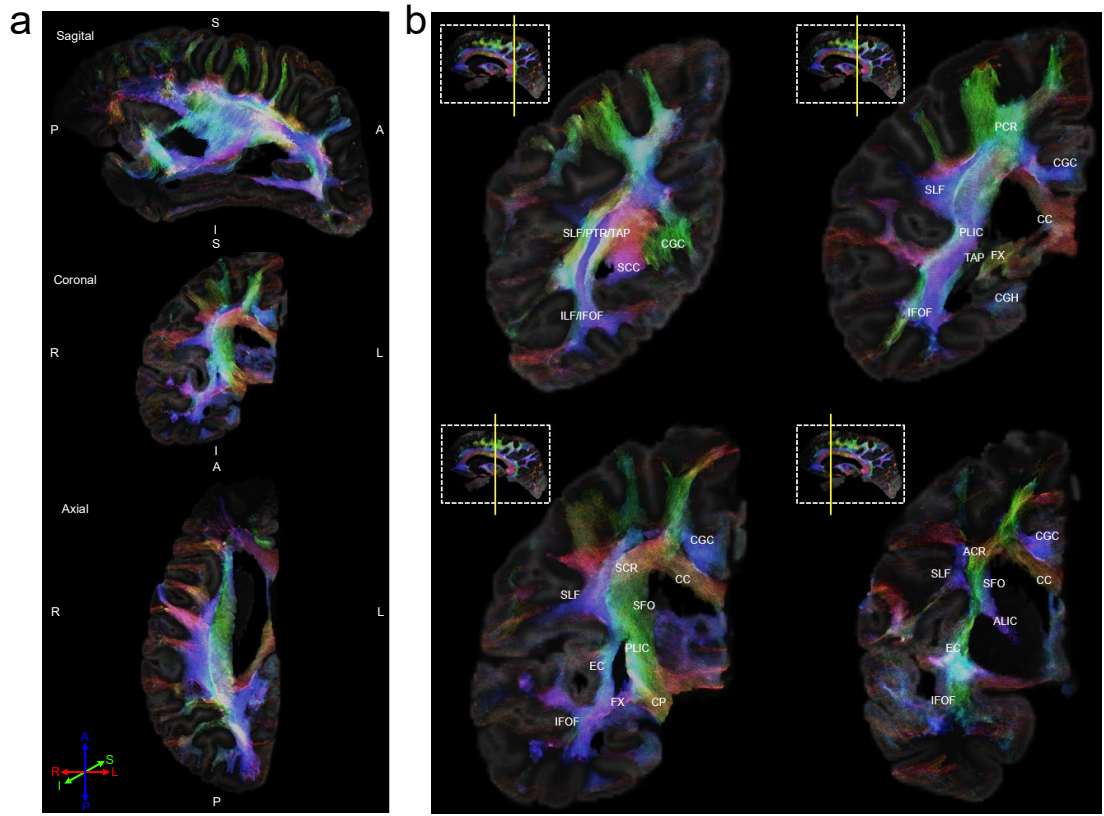

**Fig. 4** Track density imaging (TDI) of *ex-vivo* brain. (a) Overview of a hemisphere. (b) Sequential coronal slices of the TDI data with anatomical labels, according to ICBM-DTI-81 WM labels atlas [45,46]. Neuroanatomic abbreviations: CC = corpus callosum; SCC= splenium of corpus callosum; Fx = fornix; CP = cerebral peduncle; ALIC = anterior limb of internal capsule; PLIC = posterior limb of internal capsule; ACR= anterior corona radiata; SCR= superior corona radiata; PCR = posterior corona radiata; PTR = posterior thalamic radiation; ILF = Inferior longitudinal fasciculus; IFOF = inferior fronto-occipital fasciculus; EC = external capsule; CGC = cingulum; SLF = superior longitudinal fasciculus; SFO = superior fronto-occipital fasciculus; TAP = tapatum.

As an example, the corpus callosum (CC) has the largest abundance of white matter fiber tracts connecting human homologous cortical regions across two hemispheres[35], which was only faintly visible in the T2-weighted image (Fig. 5a). The colored FA map (Fig. 5b) was clearly depicted the CC, and also revealed the heterogeneous fiber orientation from the anterior to the posterior part of the CC. The same pattern can be seen on TDI (Fig. 5d) from peri-sagittal to midbrain sections. The arrangement of fibers in the CC was also visible from gross anatomy (Fig. 5c).

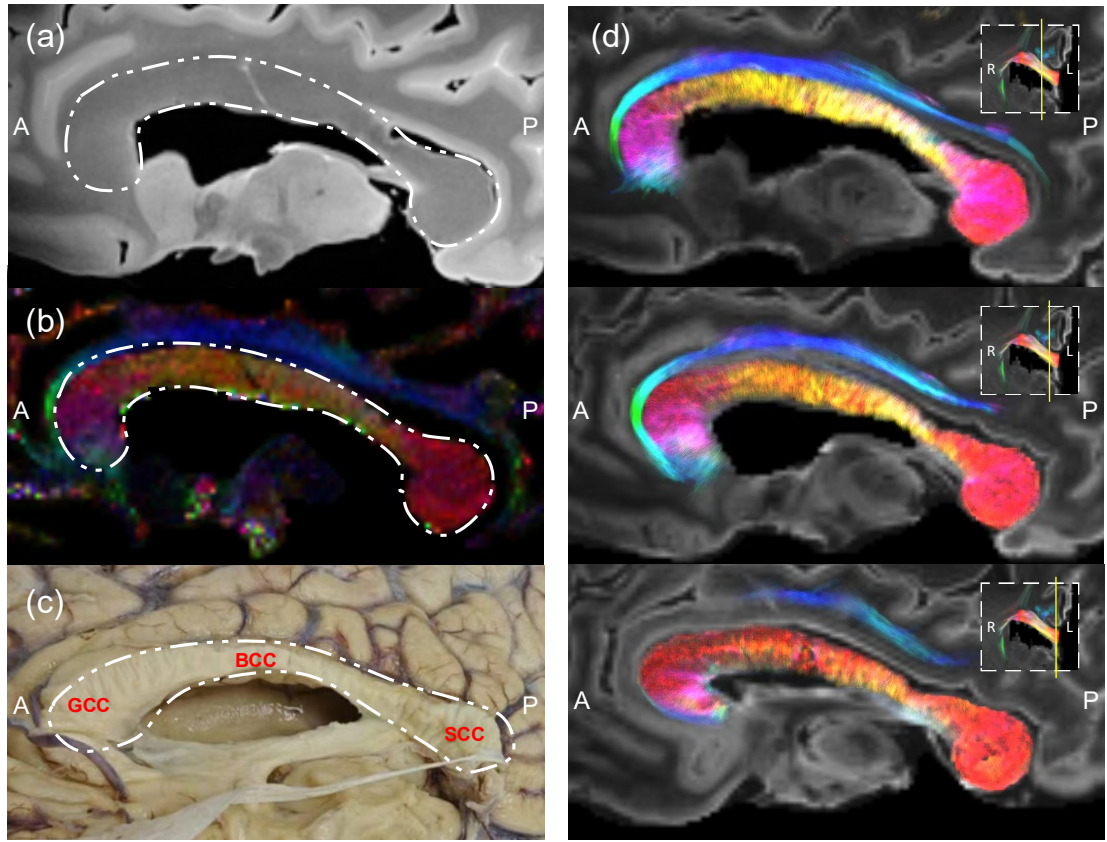

**Fig. 5** Visualization of corpus callosum (CC). (a) CC on the T2w image. (b) CC on colored FA. (c) Gross anatomy of CC. The outline of the CC is indicated by the white dashed line. (d) Sequential sagittal slices of the TDI. Neuroanatomic abbreviations: GCC = genu of corpus callosum; BCC = body of corpus callosum; SCC = splenium of corpus callosum

*Template construction*

Six hemispheres were co-registered to generate a population-average multiple modal MRI templates of the *ex-vivo* human brain. First, all *ex-vivo* hemispheres were registered to the MNI152 space [47] using affine transformations. Initially, a single subject ($S_0$) was used as the initial template and all other subjects ($S_1$, …, $S_5$) were registered to initial template, with linear and nonlinear registrations in ANTs (https://github.com/ANTsX/ANTs). After registration, all deformed brains ($S_0'$, …, $S_5'$) were averaged to create an updated template. The six brains were coregistered to the updated template and this process iterated 20 times to obtain the final template (Fig. 6). Nonlinear registration was performed using symmetric diffeomorphic deformable registration algorithm (SyN) [48], and neighborhood cross correlation was adopted as the similarity measure for images within the same modality.

Specifically, for the T2-weighted template, a dual-channel registration framework was adopted for T2w template generation using the T2w intensity images and the corresponding segmentation maps. Notably, a trained neural network used to segment the GM and WM, as described later. For the DTI template, the MD maps were used as the main contrast for image registration, as it provided clear contrast between GM and WM. Transformations estimated from the final round were applied to transform the tensor images ($D_{xx}$, $D_{yy}$, $D_{zz}$, $D_{xy}$, $D_{yz}$, $D_{xz}$) from subject space to template space. The individual tensor images were averaged to compute the population averaged DTI parametric maps. The construction of QSM templates was similar to that of T2w template, utilizing a dual-channel registration with the mean magnitude image from the 12 gradient echoes and the segmentation image. The segmentation label was obtained from the T2w image space by affine transformation. The final transformations were applied to the QSM maps to create the QSM template.

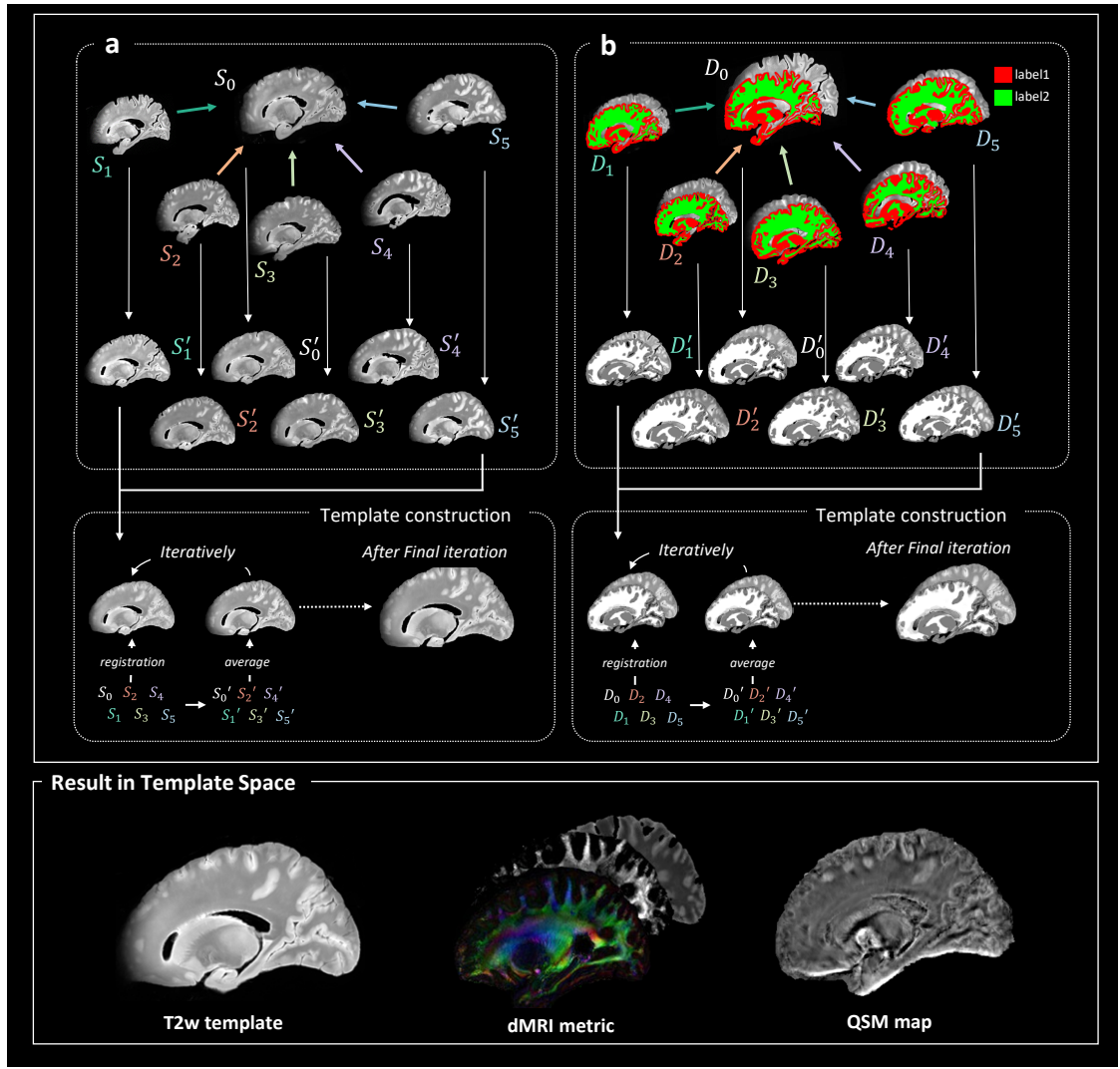

**Fig.** 6 Pipelines to generate population-average *ex-vivo* brain template for multimodal MRIs, include single-channel (a) and dual-channel registration (b).

*Anatomical labels of the ex vivo human brain*

We employed the nnU-net [49,50], a self-configuring method for automated segmentation of the cerebral cortex. The network was trained on our previously collected *ex-vivo* brain [4]. A detailed ROIs on the deep gray matter and hippocampus subregions were manually outlined by an experienced MRI physicist [Q. Zhu], with reference to the relevant literature [51,52].

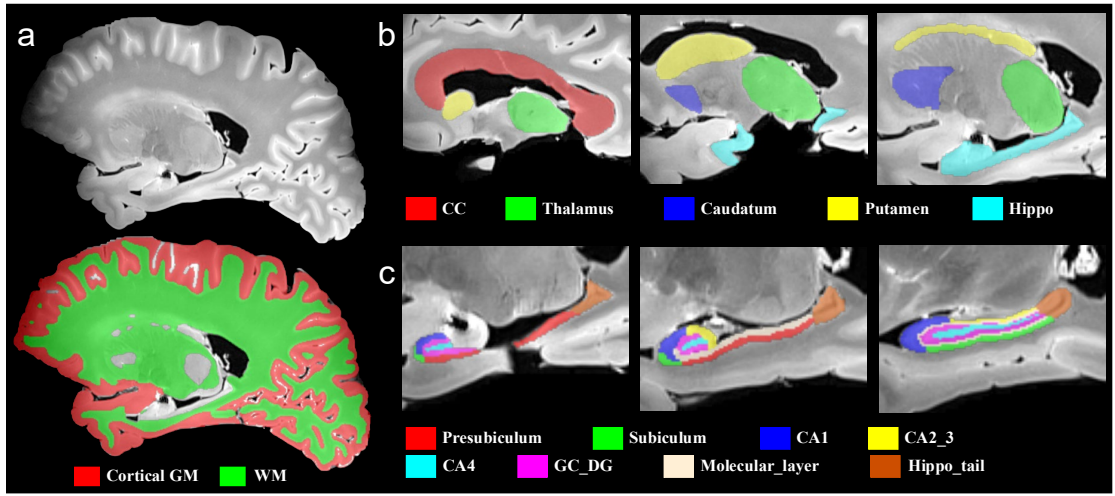

**Fig. 7** Illustration of brain region segmentation on T2w image. (a) Automated segmentation of the gray and white matter of the brain using nnU-net. (b) Manual segmentation of the deep brain structures. ROIs contained the corpus callosum (red), thalamus (green), caudatum (blue), putamen (yellow), and hippocampus (cyan). (c) Segmentation of subregions of the hippocampus. ROIs contained the presubiculum (red), subiculum (green), CA1 (blue), CA2_3 (yellow), CA4 (cyan), GC_DG (plum), molecular layer (flesh), and caudal part of the hippocampus (brown).

**Data Records**

In this study, we published all the multimodal MRI data from sample No. 4, including the raw data, the associated quantitative parametric maps, the statistical results for each metric, and the segmentation labels. We will also publish metric maps for the other five samples. In addition, we also published the multimodal template generated by co-registrating the data from all *ex-vivo* brain samples acquired, as well as the pipelines for *ex-vivo* brain alignment. All data are publicly available in figshare.

**Technical Validation**

*Assessment of spatial resolution*

We evaluated the impact of images resolutions on the accuracy of fiber tracking. Low-resolution DWI data with 1.05, 1.5, and 2 mm isotropic dimensions were obtained by downsampling the acquired high-resolution images at an isotropic resolution of 0.8 mm. Fiber tracking was performed in MRtrix 3.0, with tensor-based deterministic algorithm, with fibers length ranging from 50 to 250 mm, a step size of 0.1 mm and 500,000 fiber bundles in total. The maximum angle between successive steps was limited to 60 degrees, and the tracking was stopped when FA was less than 0.1. We tracked the ILF and IFOF by restricting the ROIs based on anatomical prior (Fig. 8a). It can be observed that the fiber bundles tracked using high-resolution data closely resemble the true anatomy, while the low-resolution dMRI can lead to false positive fiber tracking results, as shown by the red arrows (Fig. 8b).

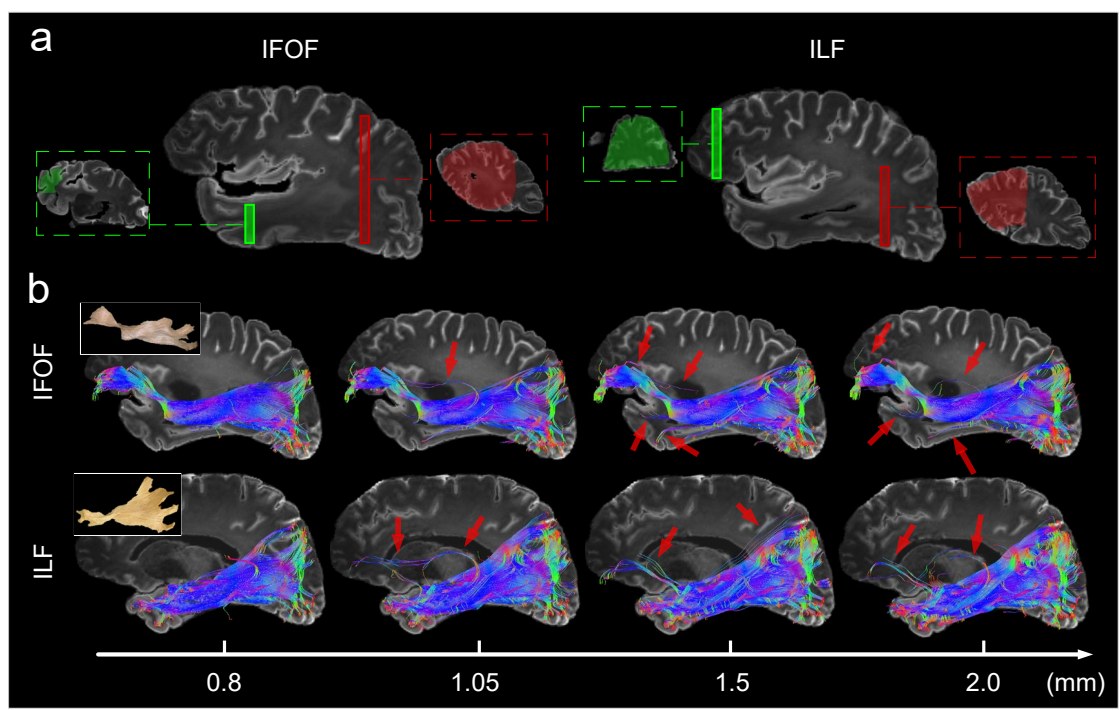

**Fig. 8** Fiber tracking of *ex-vivo* brain. (a) ROIs for filtering out fiber bundles. (b) Fiber tracking results using dMRI data at varying resolutions. False-positive results (indicated by the red arrows) in fiber tracking due to low resolution.

### *Registration mediated by the ex-vivo template*

For analysis of the *ex-vivo* MRI, we often need to register the *ex-vivo* brain to the standard *in-vivo* MRI space for segmentation or normalization, which is challenging given the large differences between the *in-vivo* and *ex-vivo* images. We compared two approaches for doing this: 1) a direct registration between the *in-vivo* and *ex-vivo* data;

and 2) utilizing the population averaged *ex-vivo* brain MRI template as an intermediate, which was already in the NMI space (Fig. 9a). Result in Fig. 9b indicated that the direct registration from *ex-vivo* to *in-vivo* template led to distortion in the ventricle, subcortex, and cortical lobes (red arrows), whereas the *ex-vivo* template-mediated approach mitigated this problem due to the similar contrast and anatomy.

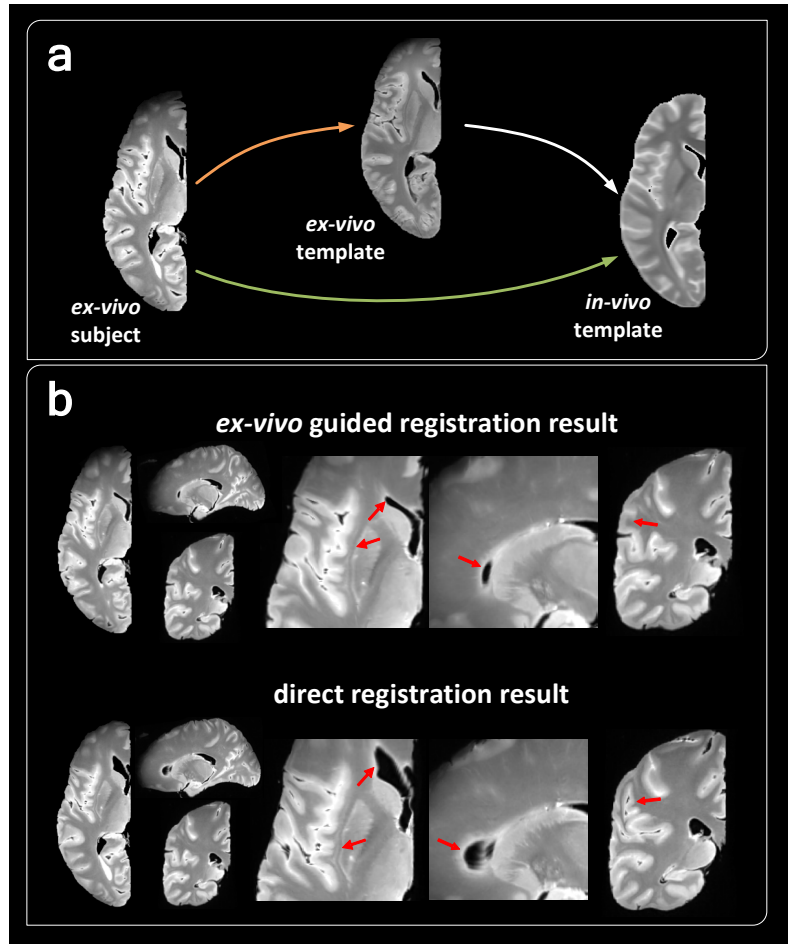

**Fig. 9** E*x-vivo* template mediated registration of *ex-vivo* MRI. (a) The direct registration approach (denoted by the green arrow) versus the *ex-vivo* template-mediated approach (noted by the orange arrow). (b) Transformed images using the two approaches, presented in axial, sagittal, and coronal views, the red arrows indicate the difference between the two registration results.

***Multi-modal quantification of the ex-vivo brains***

Based on the segmentation results, we calculated *ex-vivo* tissue properties in each brain region, in terms of the DTI metrics, DBSI metrics, FOD metrics, QSM metrics and relaxation times in Table 3. Compared to *in-vivo* data [53-55], the *ex-vivo* specimens had lower relaxation time, susceptibility, lower FA, and diffusivity in most brain regions. However, a significant decrease in diffusivity, susceptibility, T2, and T2* relaxation time was not observed in the thalamus. In DBSI, *ex-vivo* specimens exhibited smaller fiber fractions as well as larger restriction fractions than *in-vivo*[56].

**Table.3** Summary of multi-modal MRI metrics in the *ex-vivo* human brain specimens

| Parameter | Cortical GM | WM | CC | Thalamus | Caudate | Putamen | Hippo |
| --- | --- | --- | --- | --- | --- | --- | --- |
| Mean diffusivity (x$10^{-3}$ mm$^2$/s) | 0.11 | 0.04 | 0.05 | 0.08 | 0.03 | 0.06 | 0.10 |
| Axial diffusivity (x$10^{-3}$ mm$^2$/s) | 0.12 | 0.06 | 0.08 | 0.10 | 0.04 | 0.07 | 0.11 |
| Radial diffusivity (x$10^{-3}$ mm$^2$/s) | 0.10 | 0.03 | 0.04 | 0.08 | 0.03 | 0.06 | 0.09 |
| Fractional anisotropy | 0.11 | 0.34 | 0.50 | 0.17 | 0.26 | 0.17 | 0.19 |
| Fiber density | 0.74 | 0.75 | 0.96 | 0.84 | 0.24 | 0.55 | 0.89 |
| Fiber fraction (%) | 44.40 | 70.95 | 78.84 | 54.71 | 63.23 | 58.32 | 51.29 |
| Hindered isotropic fraction (%) | 24.61 | 5.35 | 6.85 | 17.13 | 7.12 | 10.68 | 21.45 |
| Restricted isotropic fraction (%) | 26.48 | 25.95 | 16.36 | 25.23 | 34.12 | 31.93 | 23.17 |
| Free isotropic fraction (%) | 7.02 | 1.56 | 2.62 | 4.51 | 2.45 | 3.06 | 6.17 |
| QSM (x$10^{-3}$ ppm) | 1.24 | -0.06 | -3.81 | 3.55 | 20.72 | 15.24 | 2.12 |
| T1 relaxation time (ms) | 1030.16 | 597.40 | 669.25 | 646.72 | 545.83 | 617.85 | 770.38 |
| T2 relaxation time (ms) | 34.84 | 25.61 | 28.43 | 27.85 | 17.17 | 22.32 | 28.96 |
| T2* relaxation time (ms) | 26.76 | 18.09 | 20.73 | 20.05 | 11.33 | 14.85 | 21.90 |

**Code availability**

The codes for parameter fitting and the weighting of segmented networks above could be accessed in figshare.

# Acknowledgements

This work was supported by the Ministry of Science and Technology of the People's Republic of China (2021ZD0200202), the National Natural Science Foundation of China (81971606, 82122032), the Science and Technology Department of Zhejiang Province (202006140, 2022C03057), the Zhejiang Provincial Natural Science Foundation of China (LQ23C090008), and the China Postdoctoral Science Foundation (GZC20241509).

# Author contributions

Consortium Leadership: Z.Z., J.Z., & D.W. Protocol Development: Q.Z., G.X., & D.W. Data collection: K.Z., Q.Z., & Z.C. Data Analysis: Q.Z., Z.C., S.L. & H.X. Code Provision: Q.Z., Z.C., & Y.S. Preparation of manuscript: Q.Z., S.L., Z.Z. & D.W.

# Competing interests

None of the authors has a conflicting financial interest.

# Additional information

Correspondence and requests for materials should be addressed to D.W.
Reprints and permissions information is available at [www.nature.com/reprints](www.nature.com/reprints).
Publisher's note Springer Nature remains neutral with regard to jurisdictional claims in published maps and institutional affiliations.